\documentclass[12pt]{article}

\usepackage{amsmath,amsfonts,amssymb}

\def\he{\hat{e}}

\def\be{\begin{equation}}

\def\ee{\end{equation}}

\def\bea{\begin{eqnarray}}

\def\eea{\end{eqnarray}}

\def\mH{\mathcal{H}}

\def\hv{\hat{v}}
\def\bz{\mathbf{z}}

\def\bG{\mathbf{G}}

\def\bx{\mathbf{x}}
\def\by{\mathbf{y}}

\newcommand{\hh}{\hat{h}}

\def\tvarphi{\tilde{\varphi}}

\newcommand{\mG}{\mathcal{G}}

\newcommand{\mL}{\mathcal{L}}

\def\pb #1{\left\{#1\right\}}

\begin{document}

\begin{titlepage}

\vskip 0.4 cm

\begin{center}
{\Large{ \bf Canonical Formalism of
        Non-Relativistic Theories Coupled to Newton-Cartan Gravity
}}

\vspace{1em}  Josef Kluso\v{n}
\footnote{Email address:
 klu@physics.muni.cz}\\
\vspace{1em} \textit{Department of Theoretical Physics and
Astrophysics, Faculty
of Science,\\
Masaryk University, Kotl\'a\v{r}sk\'a 2, 611 37, Brno, Czech Republic}\\

%
%

\vskip 0.8cm

\end{center}

\begin{abstract}
In this short note we perform canonical analysis of Schr\r{o}dinger field and
non-relativistic electrodynamics coupled to Newton-Cartan gravity. We identify physical degrees of freedom and analyze constraints structure of these theories.

\end{abstract}

\bigskip

\end{titlepage}

\newpage

\section{Introduction and Summary}\label{first}
There is a renewed interest in Newton-Cartan (NC) geometry which has
been observed in the last  years. The first significant paper was
\cite{Son:2013rqa} which introduced NC to field theory that analyzes
strongly correlated electrons. It was further shown in
\cite{Christensen:2013lma,Christensen:2013rfa} that NC geometry with
torsion naturally emerges as the background boundary geometry in
holography for $z=2$ Lifshitz geometries, for relevant works, see
\cite{Geracie:2014nka,Jensen:2014aia,Hartong:2014pma,Hartong:2015wxa}
and for review and extensive list of references, see
\cite{Taylor:2015glc}. In fact, NC geometry is non-relativistic
background geometry to which non-relativistic field theories can be
covariantly coupled, see for example
\cite{Jensen:2014aia,Geracie:2015dea,Bergshoeff:2015sic,Festuccia:2016caf,Geracie:2015xfa,Festuccia:2016awg}.
In particular, it was shown  in significant paper
\cite{Festuccia:2016caf} how non-relativistic electrodynamics can
couple to the most general NC geometry with torsion. Further,
non-relativistic scalar fields coupled to NC geometry and background
electromagnetic field were also analyzed there.

Since these results are very interesting non-relativistic theories
in NC background  certainly deserve to be studied in more details
further. In this short note we focus on canonical analysis of
non-relativistic scalar field and non-relativistic electrodynamics
coupled to NC geometry. It turns out that this is rather non-trivial
problem with interesting property that the constraints  explicit
depend on time. In more details, we start with the action for
Schr\"{o}dinger field in the background NC geometry and background
non-relativistic electromagnetic field. Such an action was derived
in \cite{Festuccia:2016caf} with the help of null reduction of
complex scalar field in higher dimensional space-time \footnote{Null
reduction was studied in some earlier papers
\cite{Duval:1984cj,Julia:1994bs}.}. Then in order to find
Hamiltonian form of this action we have to impose important
restriction on the NC space-time in the sense  that it has to have a
notion of foliation by spatial surfaces that are orthogonal to one
form $\tau_\mu$ where $\tau_\mu$ is known as clock form. This form
defines a preferred notion of spatial direction at each point and
also arrow of time in the sense that vector field $t^\mu $ is said
to be future directed if it obeys the condition $\tau_\mu t^\mu>0$.
$\tau_\mu$ defines a pointwise notion of spatial direction with the
help of the vectors $w^\mu$ that obey the condition $\tau_\mu
w^\mu=0$. However this notion can be integrated to a local
codimension one subspace when $\tau_\mu$ obeys Frobenious condition
$\tau \wedge d\tau=0$ where $\tau=\tau_\mu dx^\mu$. Then we define
causal space-times as space-times where this condition holds
everywhere, for more detailed analysis and discussion, see for
example \cite{Geracie:2015dea,Geracie:2015xfa}. For such space-time
we will be able to find Hamiltonian for the Schr\"{o}dinger field in
NC background. However we also find that when we write the complex
scalar field in polar form as $\psi=\sqrt{\rho}e^{iS}$ that the
momentum conjugate to $\rho$ is zero which is first primary
constraint of the theory. Further, in case of the momentum conjugate
to $S$ we find that it is determined by second primary constraint
that explicitly depends on time. This is very interesting situation
that deserves careful treatment. For that reason we  perform
analysis of constraint systems with explicit time dependent primary
constraints in appendix \footnote{Discussion of the constraint
analysis with explicit time dependence can be found in
\cite{Gitman:1990qh} however the analysis presented there is
slightly different from ours.}. Taking into account  explicit time
dependence of the constraints we will be able to derive canonical
equations of motion that reproduce Lagrangian equations of motion
which is a nice consistency check.

As the next step we extend this analysis to the case of non-relativistic electrodynamics in   Newton-Cartan background. Since canonical analysis is based on an existence of Lagrangian we start with the non-relativistic electrodynamics action in NC background that is derived using null dimensional reduction \cite{Festuccia:2016caf}. We again restrict to the case of causal space-time and in the first step we determine set of primary constraints which Poisson commute among themselves. This is different situation than in case of the scalar field  where the primary constraints were the second class constraints. Then the requirement of the preservation of the primary constraints gives set of secondary constraints which together with the primary constraints form set of the second class constraints. As a result we find that gauge field and corresponding conjugate momenta can be eliminated from the theory at least in principle. We also determine Lagrange multipliers corresponding to the primary constraints using the equations of motion for gauge field and we show that the resulting equations of motion coincide with the equations of motion derived by variation of action.

Let us outline main results derived in this paper. We obtain  Hamiltonian form of non-relativistic theories on NC background and we
determine physical degrees of freedom. This is very important result since we show that in case of non-relativistic electrodynamics the only physical degree of freedom is the scalar field and conjugate momenta. We also discuss the problem of the constraint structure in case of theories with explicit time dependent constraints.

The structure of this paper is as follows. In the next section (\ref{second}) we review basic facts about NC geometry and introduce an action for Schr\"{o}dinger field in the NC background and background non-relativistic electromagnetic field through null dimensional reduction. Then we perform Hamiltonian analysis of this theory and determine structure of constraints. In section (\ref{third}) we analyze non-relativistic electrodynamics in NC background. We firstly perform canonical analysis of non-relativistic electrodynamics in flat background and then we extend this analysis to the case of non-relativistic electrodynamics in NC background. Finally in appendix (\ref{Appendix})
we study  constrained systems with explicit time dependence and
discuss their properties.

\section{Hamiltonian Analysis of  Schr\"{o}dinger field in NC background}
\label{second}
\subsection{Summary of  Newton-Cartan Geometry}
We start this analysis with the brief review of Newton-Cartan geometry in $d+1$ dimensions.
Newton-Cartan background in $d+1$ dimensions is given by a set of one forms $(\tau_\mu,e_\mu^{ \ a})$ where $a=1,\dots,d$ and where $\mu,\nu=0,1,\dots,d$. We also have one form $M_\mu$. We define inverse vielbeins $v^\mu$
and $e_\mu^{ \ a}$ through the relations
\begin{equation}
v^\mu e_\mu^{ \ a}=0 \ ,  \quad v^\mu \tau_\mu=-1 \ ,
\quad e^\mu_{ \ a}\tau_\mu=0 \ , \quad e^\mu_{ \ a}e_\mu^{ \ b}=\delta_a^b \ .
\end{equation}
The determinant of the $(d+1)\times (d+1)$ matrix $(\tau_\mu,e_\mu^{ \ a})$ is denoted by $e$. With the help of  vierbeins we can construct degenerative "spatial metric"
\begin{equation}
h_{\mu\nu}=e_\mu^{ \ a}e_\nu^{ \ b}\delta_{ab} \ , \quad
h^{\mu\nu}=e^\mu_{ \ a}e^\nu_{ \ b}\delta^{ab} \ .
\end{equation}
By definitions, one forms $\tau_\mu, e_\mu^{ \ a}$ and $M_\mu$
transform under diffeomorphism as usual but they also transform
under various local transformations: Galilean boosts with
$\lambda_a$ as local parameter, local $SO(d)$ rotations which is
parameterized by $\lambda_{ab}=-\lambda_{ba}$ and $U(1)_\sigma$
gauge transformation that is parameterized by $\sigma$ where we have
\begin{eqnarray}
\delta \tau_\mu&=& 0 \ , \quad \delta e_\mu^{ \ a}=\tau_\mu \lambda^a+\lambda^a_{ \ b}e_\mu^{ \ b} \ , \nonumber \\
\delta v^\mu&=&\lambda^a e^\mu_{ \ a} \ , \quad \delta e^\mu_{ \ a}=\lambda_a^{ \ b}e^\mu_{  \ b} \ , \nonumber \\
\delta M_\mu&=&\lambda_a e_\mu^{ \ a}+\partial_\mu\sigma
 \ .\nonumber \\
 \end{eqnarray}
 The inverse vielbein $e^\mu_{ \ a}$ is invariant under local Galilean transformations.
Note that we have an important relation
\begin{equation}
e_\mu^{ \ a}e^\nu_{ \ a}-\tau_\mu v^\nu=\delta_\mu^\nu
\end{equation}
that implies
\begin{equation}
h_{\mu\nu}h^{\nu\rho}=\delta_\mu^\rho+\tau_\mu v^\rho
\end{equation}
which will be useful below. It is also useful to define  objects
that are invariant under local Galilean transformations $\hv^\mu \ ,
\he^\mu_{ \ a} \ , \hh^{\mu\nu}$ and $\Phi$ defined as
\begin{eqnarray}\label{hv}
\hv^\mu&=&v^\mu-h^{\mu\nu}M_\nu \ , \quad \he^\mu_{ \ a}=e^\mu_{ \ a}-M_\nu e^\nu_{ \ b}\delta^{ba}\tau_\mu \ ,
\nonumber \\
\hh_{\mu\nu}&=&h_{\mu\nu}-M_\mu \tau_\nu-M_\nu \tau_\mu \ ,  \quad
\Phi=-v^\mu M_\mu+\frac{1}{2}h^{\mu\nu}M_\mu M_\nu \ .
\nonumber \\
\end{eqnarray}
It is important to stress that $\hh_{\mu\nu}\neq \he_\mu^{ \ a}\he_\nu^{ \ b}\delta_{ab}$. Instead, using the definition of  $\he_\mu^{ \ a}$ given above, we obtain following relation
\begin{eqnarray}
\he_\mu^{ \ a}\he_\nu^{ \ b}\delta_{ab}=
\hh_{\mu\nu}+2\tau_\mu\tau_\nu\Phi \ . \nonumber \\
\end{eqnarray}
Finally note that  hatted objects
 obey following the relations
\begin{equation}
\hv^\mu \he_\mu^{ \ a}=0 \ , \quad \hv^\mu \tau_\mu=-1 \ , \quad
e^\mu_{ \ a}\he_\mu^{ \ b}=\delta_a^b \ .
\end{equation}
After this review of NC geometry we proceed to the Hamiltonian analysis of Schr\"{o}dinger field.
\subsection{Schr\"{o}dinger Field in NC geometry through Null Dimensional Reduction}
We would like to find Hamiltonian formulation of scalar field on Newton-Cartan background  with fixed electromagnetic background.
The most convenient way how to find such an action is to perform null dimensional reduction, see for example  \cite{Geracie:2015dea,Festuccia:2016caf}.
Let us consider  an action  for complex scalar field in
$d+2$ dimension in the form
\begin{equation}
I=\int d^{d+2}x\sqrt{-\gamma}
\left(-\gamma^{AB}D_A \Psi D_B\Psi^*\right) \ ,
\end{equation}
where $D_A=\partial_A \Psi-iq A_A\Psi$ and where $A_A, A=0,\dots,d+1$ is  background
electromagnetic field.
Let us now consider a background metric which possesses a null
isometry that is generated by coordinates $\partial_u$
\begin{equation}
ds^2=\gamma_{AB}dx^A dx^B=2\tau_\mu dx^\mu(du-M_\nu dx^\nu)+h_{\mu\nu} dx^\mu dx^\nu
\end{equation}
so that
\begin{equation}
\gamma_{\mu u}=\gamma_{u\mu}=\tau_\mu \ , \quad \gamma_{\mu\nu}=
h_{\mu\nu}-\tau_\mu M_\nu -\tau_\nu M_\nu\equiv \hh_{\mu\nu}
\ .
\end{equation}
Then
\begin{equation}
\sqrt{-\gamma}=e \ ,
e=\det \left(\tau_\mu, e_\mu^{ \ a}\right) \ .
\end{equation}
Since the metric $\gamma_{AB}$ is non-singular we can easily find inverse
metric with components
\begin{equation}\label{gammainv}
\gamma^{uu}=2\Phi \ , \quad \gamma^{u\mu}=-\hv^\mu \ , \quad  \gamma^{\mu\nu}=h^{\mu\nu} \ ,
\end{equation}
where $\hv^\mu$ and $\Phi$ are defined in (\ref{hv}).
%
%
%
We further presume that the gauge field has the form $A_A=(A_u,A_\mu)=
(\varphi,\bar{A}_\mu-\varphi M_\mu)$.
Now with the help of this metric we perform dimensional reduction of the action.
To do this we have to presume that all fields do not depend on $u$. We impose following ansatz for the scalar field $\Psi$
\begin{equation}
\Psi=e^{imu}\psi
\end{equation}
and insert it to the action. Then also using (\ref{gammainv}) we obtain
\begin{eqnarray}\label{Schroact}
I&=&\int d^{d+1}xe\left(\hv^\mu (D_\mu\psi)^*i(m-q\varphi)\psi-i(m-q\varphi)\hv^\mu D_\mu\psi
\psi^* \right.
\nonumber \\
&-& \left. 2\Phi (m-q\varphi)^2\psi \psi^*-h^{\mu\nu}D_\mu\psi (D_\nu\psi)^*\right) \ ,
\nonumber \\
\end{eqnarray}
where
\begin{equation}
D_\mu\psi=\partial_\mu\psi-iq A_\mu \psi \ , \quad (D_\mu\psi)^*=
\partial_\mu\psi^*+iq A_\mu\psi^* \ .
\end{equation}
The action (\ref{Schroact}) is the action for Schr\"{o}dinger field of the mass $m$ and charge $q$ in
Newton-Cartan background and in the background electromagnetic field
where the electromagnetic field has components $A_\mu$. Note that $\psi$ couples to $\varphi$ through the combination $m-q\varphi$ and hence
$\varphi$ effectively shifts the mass of the scalar field and hence it is natural to call it as mass potential
\cite{Festuccia:2016caf}.
 Our goal is to find Hamiltonian from the action (\ref{Schroact}).
\subsection{Hamiltonian Analysis}
We would like to work with real variables rather than with complex ones. For that
reason se introduce following parameterization of the scalar field $\psi$ as  $\psi=\sqrt{\rho}e^{iS}$ so that the action has the form
\begin{eqnarray}
I^{sch}
&=&\int d^4x e\left(2(m-q\varphi)\rho \hv^\mu \partial_\mu S-
2(m-q\varphi)\hv^\mu A_\mu\rho-2\Phi(m-q\varphi)\rho-\right.\nonumber \\
&-&\left.\frac{1}{4\rho}h^{\mu\nu}(D_\mu\rho)^*D_\nu\rho+2qh^{\mu\nu}\sqrt{\rho}
A_\mu\partial_\nu S-\rho h^{\mu\nu}\partial_\mu S\partial_\nu S\right) \ ,
\nonumber \\
\end{eqnarray}
where
\begin{equation}
D_\mu\rho=\partial_\mu\rho-2iqA_\mu\rho \ , \quad
(D_\mu\rho)^*=\partial_\mu\rho+2iqA_\mu\rho \ .
\end{equation}
 It is clear that  the previous action is well defined for  general Newton-Cartan background
for arbitrary $\tau_\mu$ apart from the fact that $\tau_\mu$ has to
obey Newton-Cartan compatibility condition. On the other hand in
order to have well defined
 Hamiltonian formulation
 we have to have a notion of  foliation by spatial surfaces that are orthogonal to $\tau_\mu$. This is guaranteed when we impose
hypersurface orthogonality condition $\tau_{[\mu}\partial_\nu
\tau_{\rho]}=0$ on the whole space-time $M$. This condition is known
as Frobenius condition and for more detailed discussion   of
causality in Newton-Cartan background, see
\cite{Geracie:2015dea,Geracie:2015xfa}. Space $M$ that obeys this
condition is called as causal. Since $\tau_\mu$ is nowhere non-zero
we can write it as $\tau_0=e^{-\Phi_L}$ where $\Phi_L$ is known as
Luttinger potential. In what follows we restrict to such space-time.
Since $\tau_i=0$ we obtain following consequences
on the form of the metric $h^{\mu\nu}$ thanks to the condition
\begin{equation}
\tau_\mu h^{\mu\nu}=0 \ .
\end{equation}
Explicitly,  for  $\nu=0$ this equation implies $\tau_0 h^{00}=0$ and hence we have to have $h^{00}=0$ while for $\nu=i$ we have
$\tau_\mu h^{\mu i}=\tau_0 h^{0i}=0$ which again implies that $h^{0i}=0$.
Then the action $I^{sch}$ simplifies considerably
\begin{eqnarray}\label{Isch}
I^{sch}&=&\int d^{d+1}x e\left(2(m-q\varphi)\rho \hv^\mu \partial_\mu S-
2(m-q\varphi)\hv^\mu A_\mu\rho-2\Phi(m-q\varphi)\rho-\right.\nonumber \\
&-& \left.\frac{1}{4\rho}h^{ij}(D_i\rho)^*D_j\rho+2qh^{ij}\sqrt{\rho}
A_i \partial_j S-\rho h^{ij}\partial_i S\partial_j S\right) \ ,
\nonumber \\
\end{eqnarray}
where $\hv^\mu=v^\mu -h^{\mu\nu}M_\nu$ has generally non-zero all its components.

Before we proceed to the Hamiltonian formulation of the theory we derive equations of motion for $\rho$ and $S$ from (\ref{Isch})
\begin{eqnarray}\label{eqIschlag}
& &\partial_\mu[e\hv^\mu (m-q\varphi)\rho]+q\partial_i[e\sqrt{\rho}h^{ij}A_j]
-\partial_i[e\rho h^{ij}\partial_jS ]
=0 \ , \nonumber \\
& &2e(m-q\varphi)\hv^\mu\partial_\mu S-2e(m-q\varphi)\hv^\mu A_\mu
-2e\Phi(m-q\varphi)+\nonumber \\
& &+\frac{e}{4\rho^2}(D_i\rho)^*D_j\rho h^{ij}
+D_i^*\left[\frac{e}{4\rho}h^{ij}D_j\rho\right]+
D_i\left[\frac{e}{4\rho}h^{ij}(D_j\rho)^*\right] \nonumber
\\
& &+
\frac{q}{\sqrt{\rho}}eh^{ij}A_i\partial_j S-eh^{ij}\partial_i S
\partial_j S=0 \ .\nonumber \\
\end{eqnarray}
 Now we are ready to proceed to the Hamiltonian formalism.
From (\ref{Isch})
 we obtain following conjugate momenta
\begin{equation}
p_S=\frac{\partial \mL^{sch}}{\partial(\partial_t S)}=2(m-q\varphi)\rho \hv^0 \ , \quad p_\rho=\frac{\partial \mL^{sch}}{\partial(\partial_t \rho)}= 0 \ .
\end{equation}
From these two equations we see that there are two primary constraints
\begin{eqnarray}
\mG_S\equiv p_S-2e(m-q\varphi)\rho\hv^0 \approx 0 \ , \quad
\mG_\rho\equiv p_\rho\approx 0 \  \nonumber \\
\end{eqnarray}
while the bare Hamiltonian  is equal to
\begin{eqnarray}
H_B&=&\int d^d\bx (p_\rho\partial_t\rho+p_S\partial_t S-\mL)=
\int d^d\bx\mH_B\ , \nonumber \\
\mH_B&=&-2e(m-q\varphi)\rho\hv^i \partial_iS+2e(m-q\varphi)\hv^\mu A_\mu\rho
+2e\Phi(m-q\varphi)\rho+\nonumber \\
&+&\frac{1}{4\rho}eh^{ij}
(D_i\rho)^*D_j\rho-2qeh^{ij}\sqrt{\rho}A_i \partial_j S+
\rho e h^{ij}\partial_i S\partial_j S \ .
\nonumber \\
\end{eqnarray}
We see that generally $\mG_S\approx 0 $ and $\mH_B$  explicit depend on time. This is not usual situation and we discuss theory of constraints systems with explicit time dependence in more details in Appendix (\ref{Appendix}).

As the next step we calculate Poisson bracket between
 $\mG_S$ and $\mG_\rho$ and we obtain
\begin{equation}
\pb{\mG_S(\bx),\mG_\rho(\by)}=-2e(m-q\varphi)\hv^0\delta(\bx-\by)\equiv
\triangle_{S\rho}(\bx,\by)
\end{equation}
which show that they are two second class constraints. Note that the inverse matrix has the form  $\triangle^{\rho S}=-\frac{1}{2e(m-q\varphi)\hv^ 0}\delta(\bx-\by)$.
As a result we can eliminate $p_\rho=0$ and $p_S=0$ from the set of canonical variables when we introduce Dirac bracket between $\rho$ and
$S$ defined as
\begin{eqnarray}\label{rhoSdir}
& &\pb{\rho(\bx),S(\by)}_D=\pb{\rho(\bx),S(\by)}
-\int  d^d\bz d^d\bz'\pb{\rho(\bx),\mG_S(\bz)}\triangle^{S\rho}(\bz,\bz')
\pb{p_\rho(\bz'),S(\by)}-\nonumber \\
&-&\int  d^d\bz d^d\bz'\pb{\rho(\bx),p_\rho(\bz)}\triangle^{\rho S}(\bz,\bz')
\pb{\mG_S(\bz'),S(\by)}
=-\frac{1}{2e\hv^0(m-q\varphi)}\delta(\bx-\by) \ . \nonumber \\
\end{eqnarray}
As was explicitly shown in Appendix (\ref{Appendix}),  in the presence of the
time dependent constraints the equations of motion for canonical variables have
the form \footnote{See equation  (\ref{eqtimecon}) in Appendix (\ref{Appendix}).}
\begin{eqnarray}
\partial_t \rho&=&\pb{\rho,H_B}_D-\nonumber \\
&-&\int d^d\bz d^d\bz' \pb{\rho,\mG_S(\bz)}\triangle^{S\rho}(\bz,\bz')\frac{\partial p_\rho}{\partial t}-
\int d^d\bz d^d\bz' \pb{\rho,p_\rho(\bz)}\triangle^{\rho S}(\bz,\bz')\frac{\partial \mG_S(\bz')}{\partial t}=
\nonumber \\
&=&
-\frac{1}{e\hv^ 0(m-q\varphi)}\partial_i[(m-q\varphi )e\hv^i\rho]-
\frac{1}{e\hv^0(m-q\varphi)}\partial_t[e(m-q\varphi)\hv^0]\rho- \nonumber \\
&-&\frac{q}{e\hv^0(m-q\varphi)}\partial_i[e\sqrt{\rho}h^{ij}A_j]
+\frac{1}{e\hv^0(m-q\varphi)}\partial_i[eh^{ij}\rho\partial_j S]  \nonumber \\
\end{eqnarray}
that can be rewritten into more symmetric form
\begin{eqnarray}
\partial_t [ e\hv^0(m-q\varphi)\rho]+\partial_i[e\hv^i(m-q\varphi)\rho]
+\partial_i[e\sqrt{\rho}h^{ij}A_j]-
\partial_i[eh^{ij}\rho\partial_j S] =0
\nonumber \\
\end{eqnarray}
that coincides with the first equation of motion  given in (\ref{eqIschlag}). Let us now proceed to the canonical equation of motion for
$S$
\begin{eqnarray}
& &\partial_t S=\pb{S,H_B}_D-
\int d^d\bz d^d\bz'\pb{S,\mG_S(
    \bz)}\triangle^{S\rho}(\bz,\bz')\frac{\partial \mG_\rho}{\partial t}
\nonumber \\
&=&- \frac{\hv^i}{\hv^0}\partial_i S+\frac{\hv^\mu}{\hv^0} A_\mu+\frac{1}{\hv^0}\Phi-\frac{1}{8\rho^2\hv^0(m-q\varphi)}h^{ij}(D_i\rho)^*D_j\rho
\nonumber \\
&-&\frac{1}{2e\hv^0(m-q\varphi)}D_i\left[\frac{eh^{ij}}{4\rho}D_j\rho\right]
-\frac{1}{2e\hv^0(m-q\varphi)}D^*_i\left[\frac{e h^{ij}}{4\rho}D^*_j\rho\right]
\nonumber \\
&-&\frac{eq}{2\hv^0(m-q\varphi)\sqrt{\rho}}h^{ij}A_i\partial_jS
+\frac{e}{\hv^0(m-q\varphi)}h^{ij}\partial_i S\partial_j S
\nonumber \\
\end{eqnarray}
that can be again rewritten into the form
\begin{eqnarray}
& &2e(m-q\varphi)\hv^\mu\partial_\mu S-2e(m-q\varphi)\hv^\mu A_\mu-
2e(m-q\varphi)\Phi+\nonumber \\
& &+\frac{e}{4\rho^2}h^{ij}(D_i\rho)^*D_j\rho+
D_i\left[\frac{e}{4\rho}h^{ij}D_j^*\rho\right]+
D_i^*\left[\frac{e}{4\rho}h^{ij}D_j\rho \right]
+\nonumber \\
& &+\frac{qe}{\sqrt{\rho}}h^{ij}A_i\partial_j S
-eh^{ij}\partial_i S\partial_j S=0
\nonumber \\
\end{eqnarray}
that coincides with the second equation of motion given in
(\ref{eqIschlag}).

In summary, we found the Hamiltonian formulation of  Schr\"{o}dinger field in NC background.
We found that the dynamical fields are $\rho$ and $S$ that have non-zero Dirac bracket
(\ref{rhoSdir}). Then we derived their canonical equations of motion and found that they coincide with the equations of motion derived from Lagrangian.

\section{Hamiltonian Formalism for Electromagnetic Field in
Newton-Cartan Gravity}\label{third}
In this section we focus on canonical analysis of non-relativistic electromagnetic
field in NC background. We start with the simpler case of the action for non-relativistic electrodynamics in flat background.
\subsection{Non-Relativistic Electrodynamics through Null Dimensional Reduction}
Following   \cite{Festuccia:2016caf}
we derive an action for non-relativistic electrodynamics by
 performing a null reduction of the Maxwell
action in one higher dimension. More precisely, let us consider $d+2$ dimensional
Maxwell action
\begin{equation}
S=-\frac{1}{4}\int dt du d^d\bx F_{AB}\eta^{AC}\eta^{BD}F_{CD}
\ ,
\end{equation}
where $\eta_{AB}dx^Adx^B=2dtdu+dx^idx^i$. Following     \cite{Festuccia:2016caf}
we set $A_u=\varphi \ , A_t=-\tvarphi \ ,  A_i=a_i$ and presume that all fields
do not depend on $u$. Since the inverse metric has the form $\eta^{tu}=\eta^{ut}=1 \ ,
\eta^{ij}=\delta^{ij}$ we get
\begin{equation}
F_{AB}\eta^{AC}\eta^{BD}F_{CD}=-2(F_{tu})^2+F_{ij}F^{ij}-4F_{iu}F_{tk}=
-2(\partial_t\varphi)^2-4(\partial_t a_i+\partial_i\tvarphi)\partial_i\varphi
+f_{ij}F^{ij} \ .
\end{equation}
As a result we obtain an action for non-relativistic electrodynamics in flat
background in the form
\begin{equation}\label{actnonEM}
S=\int dt d^d\bx\left(-\frac{1}{4}f_{ij}f^{ij}+(\partial_i\tvarphi+\partial_t a_i)\partial_i\varphi+\frac{1}{2}(\partial_t \varphi)^2\right) \ ,
\end{equation}
where $f_{ij}=\partial_i a_j-\partial_j a_i$.  From the action (\ref{actnonEM}) we
derive following conjugate momenta
\footnote{In this section we do not carry about upper or lower spatial index since they are equivalent in flat background.}
\begin{equation}
\pi^i=\partial_i \varphi \  , \quad p_{\tvarphi}= 0 \ , \quad
p_\varphi=\partial_t \varphi
\end{equation}
so that we have following primary constraints
\begin{eqnarray}\label{primnonEM}
\mG^i\equiv
\pi^i-\partial_i\varphi \approx 0 \ , \quad  p_{\tvarphi}\approx 0 \ ,
\nonumber \\
\end{eqnarray}
together with the bare Hamiltonian in the form
\begin{equation}
H_B=\int d^d\bx (\pi^i\partial_t a_i
+p_\varphi\partial_t\varphi+p_{\tvarphi}\partial_t \tvarphi-\mL)
=
\int d^d\bx\left(\frac{1}{4}f_{ij}f^{ij}-\partial_i \tvarphi \partial_i\varphi
+\frac{1}{2}p_\varphi^2\right) \
\end{equation}
and consequently the extended Hamiltonian is equal to
\begin{equation}
H_E=H_B+\int d^d\bx(\lambda_i\mG^i+\lambda^{\tvarphi} p_{\tvarphi}) \ ,
\end{equation}
where $\lambda_i,\lambda^{\tvarphi}$ are Lagrange multipliers corresponding to the constraints $\mG^i\approx 0$ and $p_{\tvarphi}\approx 0$.
As the next step we have to ensure the preservation of all primary constraints.
In case of the constraint $\mG^i\approx 0$ we obtain
\begin{equation}
\frac{d\mG^i}{dt}=\pb{\mG^i,H_E}=\partial_k f^{ki}-\partial_i p_\phi \equiv \mG^i_{II}\approx 0 \ ,
\end{equation}
where $\mG^i_{II}\approx 0$ are secondary constraints. In case of the constraint
 $p_{\tvarphi}\approx 0$ we obtain
\begin{equation}
\frac{dp_{\tvarphi}}{dt}=\pb{p_{\tvarphi},H_E}=\partial_i \partial^i\varphi
=\partial_i\pi^i-\partial_i \mG^i\approx \partial_i\pi^i\equiv
\mG_{\tvarphi}^{II}\approx 0 \
\end{equation}
which is the generator of gauge transformations. In fact, if we define
\begin{equation}
\bG(\Lambda)=\int d^d \bx \Lambda \mG^{II}_{\tvarphi}
\end{equation}
we obtain standard transformation rules
\begin{equation}
\pb{\bG(\Lambda),A_i}=\partial_i \Lambda \ , \quad
\pb{\bG(\Lambda),f_{ij}}=0 \ .
\end{equation}
Finally we have to ensure the preservation of the constraint $\mG^i_{II}\approx 0$.
To do this we have to calculate the Poisson bracket between constraints
 $\mG^i$ and $\mG^i_{II}$. After some calculations we obtain
\begin{eqnarray}\label{triangleij}
\pb{\mG^i(\bx),\mG^j_{II}(\by)}=
-\partial_k\partial^k\delta(\bx-\by)\delta^{ij}\equiv
\triangle^{ij}(\bx,\by) \ .
\nonumber \\
\end{eqnarray}
Let us introduce an inverse matrix $D_{ij}(\bx,\by)$
that obeys the relation
\begin{equation}
\int d^d \bz \triangle^{ik}(\bx,\bz)D_{kj}(\bz,\by)=\delta^i_j\delta(\bx-\by) \ .
\end{equation}
Since $\triangle_{ij}$ is given in (\ref{triangleij}) we find that
 $D_{ij}$ is a solution of the
equation
\begin{equation}
\frac{\partial}{\partial x^k}\frac{\partial}{\partial x^k}D_{ij}(\bx,\by)=-
\delta_{ij}\delta(\bx-\by) \ .
\end{equation}
As the next step we determine  canonical equations of motion for $\varphi$ and $p_\varphi$
\begin{eqnarray}\label{eqvarphi}
\partial_t\varphi&=&\pb{\tvarphi,H_E}=
\pb{\tvarphi,H_B}+\pb{\varphi,\int d^d\bz \lambda^i\mG_i(\bz)}=p_\varphi
\nonumber \\
\partial_t p_{\varphi}&=&\pb{p_{\varphi},H_E}=
\pb{p_{\varphi},H_B}+\pb{p_{\tvarphi},\int d^d\bz \lambda^i\mG_i(\bz)}=
-\partial_k\partial_k \tvarphi-\partial_i \lambda^i \ .
\nonumber \\
\end{eqnarray}
Finally the equation of motion for $a_i$ has the form
\begin{equation}
\partial_t a_i=\pb{a_i,H_E}=\int d^d\bz \lambda_j(\bz)\pb{a_i,\mG^j(\bz)}=
\lambda_i
\end{equation}
so that the equation of motion for $p_\varphi$ can be written as
\begin{equation}
\partial_t p_{\varphi}=-\partial_k(\partial_k\tvarphi+\partial_t a_k)=\partial_k
\tilde{E}_k \ ,
\end{equation}
where $\tilde{E}_k=-\partial_t a_k-\partial_k\tvarphi$. If we perform  partial
time derivation  of the first  equation in (\ref{eqvarphi})  and use the second one we obtain
\begin{equation}
\partial_t^2 \varphi=\partial_t p_{\varphi}=\partial_k \tilde{E}_k
\end{equation}
with agreement with the equation (2.11) in \cite{Festuccia:2016caf}.  Further,
if we apply  the partial derivative $\partial_i$ on the first equation in
(\ref{eqvarphi}) we obtain
\begin{equation}
\partial_i\partial_t \varphi=\partial_ip_{\varphi}=\partial_k f^{ki}
\end{equation}
with agree with the second equation in (2.3) in \cite{Festuccia:2016caf}. In the same way we find that the divergence of $\mG^i=0$ implies
\begin{equation}
\partial_i \mG^i=
\mG^{II}_{\tvarphi}-\partial_i\partial^i\varphi \approx
- \partial_i\partial^i\tvarphi=0
\end{equation}
that agrees with the first equation in (2.3)
\cite{Festuccia:2016caf}.
Finally we should determine the Lagrange multiplier $\lambda_i$ using the requirement of the preservation of the constraint $\mG^i_{II}$ but this is not necessary since we know that $\lambda_i=\partial_t a_i$. On the other hand  since $\mG^i\approx 0$ and $\mG^i_{II}\approx 0$ are two second class constraints they can be explicitly solved for $\pi^i$ and $a_i$. In other words there is only one dynamical variable which is $\varphi$ and its conjugate momentum $p_{\varphi}$.

\subsection{Null Reduction of Maxwellian Electromagnetism in NC Background}
We determine action for electromagnetic field in Newton-Cartan background
again with the help of null dimensional reduction, following
\cite{Festuccia:2016caf}. We start with the action
for electromagnetic field in $d+2$ dimensions that has the form
\begin{equation}
S=-\frac{1}{4}\int d^{d+2}x\sqrt{-\gamma}F_{AB}\gamma^{AC}\gamma^{BD}F_{CD} \ .
\end{equation}
Our goal is to dimensionally reduce this action along null isometry so that we will presume that $A_M$  do not depend on $u$. We further write $A_M=(A_u,A_\mu)$ and define $A_\mu\equiv \varphi$.
Since the gauge field transforms under $U(1)$ transformations as
\begin{equation}
A'_A=A_A+\partial_A \Lambda
\end{equation}
it is clear that $\varphi$ is invariant under gauge transformation since $\Lambda$ does not depend on $u$. On the other hand the gauge field $A_\mu$ transform as
\begin{equation}
A_\mu'=A_\mu+\partial_\mu \Lambda \ .
\end{equation}
In order to perform null dimensional reductions we use the components of  metric inverse given in (\ref{gammainv}) and we obtain the action in the form
\begin{equation}\label{SelectroNCi}
S=\int d^{d+1}x
e \left(-\frac{1}{4}F_{\mu\nu}h^{\mu\rho}h^{\nu\sigma}F_{\rho\sigma}-
\Phi \partial_\mu \varphi h^{\mu\nu}\partial_\nu\varphi+
\frac{1}{2}(\hv^\mu\partial_\mu\varphi)^2-\hv^\nu F_{\nu\mu}h^{\mu\sigma}
\partial_\sigma\varphi\right) \ ,
\end{equation}
where
\begin{equation}
F_{\mu\nu}=\partial_\mu A_\nu-\partial_\nu A_\mu \ .
\end{equation}
It is convenient to use slightly different form of the action which
depends on $v^\mu$ instead of $\hv^\mu$. Following
\cite{Festuccia:2016caf}
we introduce vector field $\bar{A}_\mu$ defined as
\begin{equation}
A_\mu=\bar{A}_\mu-\varphi M_\mu \ .
\end{equation}
Performing this substitution in the action (\ref{SelectroNCi}) we find
\begin{equation}\label{SelectroNC}
S=\int d^{d+1}x e\left(-\frac{1}{4}h^{\mu\rho}h^{\nu\sigma}\bar{F}_{\mu\nu}
\bar{F}_{\rho\sigma}-h^{\mu\nu}v^\rho
\bar{F}_{\rho\nu}\partial_\mu\varphi+\frac{1}{2}(v^\mu \partial_\mu\varphi)^2\right)
\ ,
\end{equation}
where
\begin{equation}
\bar{F}_{\mu\nu}=\partial_\mu \bar{A}_\nu-\partial_\nu\bar{A}_\nu-
\varphi(\partial_\mu M_\nu-\partial_\nu M_\mu) \ .
\end{equation}
The  action (\ref{SelectroNC})
 will be the starting point for the Hamiltonian formulation of the theory. As we argued above we restrict to causal space-time with non-zero $\tau_0$ only. Then $v^\mu$ has generally non-zero all components with $v^0=-\tau_0$.
In case of causal space-time the action has the form
\begin{equation}\label{actNCelcaus}
S=\int d^{d+1}xe\left(-\frac{1}{4}\bar{F}_{ij}h^{ik}h^{jl}\bar{F}_{kl}-
\Phi \partial_i\varphi h^{ij}\partial_j\varphi+\frac{1}{2}
(v^\mu\partial_\mu\varphi)^2-v^\nu \bar{F}_{\nu i}h^{ij}\partial_j\varphi\right) \ .
\end{equation}
Note that from this action we also obtain equations of motion in the form
\begin{eqnarray}\label{eqelectroNC}
& &2\partial_i[e\Phi h^{ij}\partial_j\varphi]-\partial_\mu[ev^\mu v^\nu\partial_\nu\varphi]+
\partial_j[ev^\nu \bar{F}_{\nu i}h^{ij}]+\nonumber \\
&+&
\frac{1}{2}e(\partial_i M_j-\partial_j M_i)h^{ik}h^{jl}\bar{F}_{kl}
+ev^0(\partial_0 M_i-\partial_i M_0)h^{ij}\partial_j\varphi+
ev^k(\partial_k M_i-\partial_i M_k)h^{ij}\partial_j\varphi
=0 \ , \nonumber \\
& &\partial_j[eh^{il}\bar{F}_{lk}h^{kj}]+\partial_0[ev^0 h^{ij}\partial_j\varphi]
+\partial_k[e v^k h^{ij}\partial_j\varphi]-\partial_k[e v^i h^{kj}\partial_j\varphi]=0 \ , \nonumber \\
& & \partial_i[ev^0 h^{ij}\partial_j\varphi]=0 \ . \nonumber \\
\end{eqnarray}
Let us now proceed to the canonical analysis.
From (\ref{actNCelcaus}) we obtain following
 conjugate momenta
\begin{eqnarray}\label{momenta}
\pi^i&=&\frac{\partial \mL}{\partial (\partial_t \bar{A}_i)}=-ev^0 h^{ij}\partial_j\varphi
 \ , \quad  \pi^0=\frac{\partial \mL}{\partial (\partial_t \bar{A}_0)}\approx 0 \ , \nonumber \\
p_\varphi&=&\frac{\partial \mL}{\partial (\partial_t \varphi)}=ev^0(v^\mu\partial_\mu\varphi)
\nonumber \\
\end{eqnarray}
so that we have following explicitly time dependent primary constraints
\begin{equation}
\mG^i\equiv \pi^i+ev^0 h^{ij}\partial_j\varphi \approx 0
\end{equation}
together with familiar constraint $\mG^0\equiv\pi^0\approx 0$. Further, with the help of
 (\ref{momenta}), we obtain  the bare Hamiltonian in the form
\begin{eqnarray}
H_B
&=&\int d^d\bx \left(e\frac{1}{4}\bar{F}_{ij}h^{ik}h^{jl}\bar{F}_{kl}+e\Phi \partial_i\varphi
h^{ij}\partial_j\varphi+ev^k \bar{F}_{ki}h^{ij}\partial_j\varphi+\right.\nonumber \\
&+& \left.\frac{1}{2e}(\tau_0)^2
p_\varphi^2+\tau_0 v^i\partial_i\varphi
 p_\varphi-A_0\partial_i\pi^i+\pi^i
 \varphi(\partial_0 M_i-\partial_i M_0)\right) \ .  \nonumber \\
\end{eqnarray}
Now we have to analyze the requirement of the preservation of primary constraints
$\mG^i\approx 0 \ , \mG^0\approx 0$. Note that the extended Hamiltonian has the form
\begin{equation}
H_E=H_B+\int d^d\bx (\lambda_i\mG^i+\lambda_0\pi^0) \ .
\end{equation}
 In case of $\mG^0\approx 0$ we obtain that the requirement of its preservation during the time development of the system implies standard
Gauss law constraint
\begin{equation}
\mG^{II}\equiv \partial_i\pi^i\approx 0 \ ,
\end{equation}
while in case of $\mG^i$ we get
\begin{eqnarray}
\frac{d\mG^i}{dt}&=&\frac{\partial \mG^i}{\partial t}
+\pb{\mG^i,H_E}\nonumber \\
&=&\partial_t (ev^0 h^{ij})\partial_j\varphi
-\partial_k [e h^{ik}\bar{F}_{kl}h^{lj}]+\partial_k(e v^k h^{in}\partial_n\varphi)-
\nonumber \\
&-&\partial_m (ev^i h^{mn}\partial_n \varphi)+ev^0 h^{ij}\partial_j [\frac{1}{e} (\tau_0)^2p_{\varphi}]+
ev^0 h^{ij}\partial_j [\tau_0 v^m\partial_m\varphi] \equiv \mG_{II}^i\approx 0 \ ,
\nonumber \\
\end{eqnarray}
where we used the fact that $\pb{\mG^i(\bx),\mG^j(\by)}=0$. We see that the requirement of the preservation of the constraints $\mG^i\approx 0$ implies the second set of the constraints $\mG^i_{II}\approx 0$.
It is again easy to see that $\mG^i\approx 0 \ , \mG^j_{II}\approx 0$ are two sets of second class constraints with rather complicated Poisson bracket between them. Then it is difficult to determine Lagrange multipliers $\lambda_i$
from the requirement of the preservation of the constraints $\mG^i_{II}\approx 0$ during the time evolution of the system. On the other hand, as we will show below, these Lagrange multipliers can be determined with the help of the equations of motion for $\bar{A}_i$.  Further, it is easy to see that $\mG^0$ and $\mG^{II}\approx 0$ are first class constraints where $\mG^{II}\approx 0$ is generator of gauge transformations.

As we argued above $\mG^i\approx 0 \ , \mG^i_{II}\approx 0$ are two sets of second
class constraints where $\mG^i=0$ can be solved for $\pi^i$ while  $\mG^i_{II}=0$ can be solved for $\bar{A}_i$ at least in principle. On the other hand when we try to write equations of motion for $\varphi$ and $p_{\varphi}$ it is convenient to express Lagrange multiplier $\lambda_i$ as a function of  non-dynamical variable $\bar{A}_i$ using its equation of motion
\begin{equation}
\partial_t \bar{A}_i=\pb{\bar{A}_i,H_E}=
\partial_i A_0+\lambda_i+\varphi (\partial_0 M_i-\partial_i M_0)
\end{equation}
that implies that $\lambda_i=\bar{F}_{0i}$.
Then we can  write canonical equations of motion for
$\varphi$ and $p_{\varphi}$ as
 \begin{eqnarray}
 \partial_t \varphi&=&\pb{\varphi,H_B}+\int d^d\bz \lambda_i\pb{\varphi,\mG^i(\bz)}=
 \frac{1}{e(v^0)^2}p_{\varphi}+\tau_0 v^i\partial_i\varphi \ , \nonumber \\
 \partial_t p_{\varphi}&=&\pb{p_{\varphi},H_B}
 +\int d^d\bz \lambda_i(\bz)\pb{p_{\varphi},\mG^i(\bz)}
 =\nonumber \\
 &=&\frac{1}{2}e
 (\partial_i M_j-\partial_jM_i)h^{ik}h^{jl}\bar{F}_{kl}
 -\pi^i(\partial_0 M_i-\partial_i M_0)+\partial_m (\lambda_j h^{jm}e)+\nonumber \\
&+&2\partial_i[e\Phi h^{ij}\partial_j\varphi]+
 \partial_j[ev^k\bar{F}_{ki}h^{ij}]+\partial_i[\tau_0 v^ip_{\varphi}]
 +ev^k (\partial_k M_i-\partial_i M_k)h^{ij}\partial_j\varphi
 \nonumber \\
 &\approx&
 \frac{1}{2}e
 (\partial_i M_j-\partial_jM_i)h^{ik}h^{jl}\bar{F}_{kl}
 +v^0eh^{ij}\partial_j\varphi (\partial_0 M_i-\partial_i M_0)+
 \partial_m (\bar{F}_{0j}h^{jm}ev^0)+
 \nonumber \\
&+&2\partial_i[e\Phi h^{ij}\partial_j\varphi]+
 \partial_j[ev^k\bar{F}_{ki}h^{ij}]+\partial_i[\tau_0 v^ip_{\varphi}]
 +ev^k (\partial_k M_i-\partial_i M_k)h^{ij}\partial_j\varphi \ .
 \nonumber \\
 \end{eqnarray}
 If we combine these two equations together we obtain
 \begin{eqnarray}
& & \partial_t (e(v^0)^2\partial_t\varphi)=\partial_t p_{\varphi}-\partial_t(ev^0v^i\partial_i\varphi)\nonumber \\
& & =\frac{1}{2}e
 (\partial_i M_j-\partial_jM_i)h^{ik}h^{jl}\bar{F}_{kl}
 +ev^0h^{ij}\partial_j\varphi(\partial_0 M_i-\partial_i M_0)+\partial_m (\bar{F}_{0j}h^{jm}ev^0)
 \nonumber \\
& & +2\partial_i[e\Phi h^{ij}\partial_j\varphi]+
 \partial_j[ev^k\bar{F}_{ki}h^{ij}]-\partial_i[e  v^i v^0\partial_t\varphi]
 +ev^k (\partial_k M_i-\partial_i M_k)h^{ij}\partial_j\varphi \
 \nonumber \\
 \end{eqnarray}
 that coincides with the first equation in (\ref{eqelectroNC}). Further, it is easy to see that the second equation in (\ref{eqelectroNC}) coincides with the secondary constraint $\mG^i_{II}=0$. Finally, the last equation in (\ref{eqelectroNC}) is equivalent to the combination of the primary constraint $\mG^{II}$ and $\mG^i$ since
\begin{equation}
\mG^{II}=\partial_i [\mG^i-ev^0 h^{ij}\partial_j\varphi]=
-\partial_i[ev^0 h^{ij}\partial_j\varphi]=0 \ .
\end{equation}
In summary, we have shown that canonical equations of motion and constraints reproduce Lagrangian equations of motion. We have also determined physical degrees of freedom of non-relativistic electrodynamics and we have shown that there are only two phase space  physical degrees left corresponding to $\varphi$ and $p_\varphi$.

 {\bf Acknowledgements:}

I would like to thank M. Chaichian for his comments and remarks. This  work  was
    supported by the Grant Agency of the Czech Republic under the grant
    P201/12/G028.

 \appendix

 \section{Appendix: Systems with explicit time dependent constraints}\label{Appendix}

 Let us consider phase space system with variables $p_m,q^m, m=1,\dots,N$, bare Hamiltonian $H_B$ and set of primary constraints $\phi_j=\phi_j(p,q,t), j=1,\dots,J$, that explicitly depend on time $t$. Then the phase space action with primary constraints included has the form
 \begin{equation}
 S=\int dt (p_m\dot{q}^ m-H_B-\lambda^j \phi_j) \ ,
 \end{equation}
 where $\lambda^j$ are independent variables known as Lagrange multipliers. Variation of the action with respect to $p_m,q^m$ and $\lambda^j$ we obtain following set of equations of motion
 \begin{eqnarray}
 & &\dot{q}^m-\frac{\partial H_B}{\partial p_m}-\lambda^j\frac{\partial \phi_j}{
    \partial p_m}=0 \ , \nonumber \\
 & &-\dot{p}_m-\frac{\partial H_B}{\partial q^m}-\lambda^j\frac{\partial \phi_j}
 {\partial q^m}=0 \ ,  \nonumber \\
 & &\phi_j\approx 0 \ . \nonumber \\
 \end{eqnarray}
 Introducing standard Poisson bracket they can be written as
 \begin{eqnarray}
 & &\dot{q}^m=\pb{q^m,H_B}+\lambda^j\pb{q^m,\phi_j} \ , \nonumber \\
 & &\dot{p}_m=\pb{p_m,H_B}+\lambda^ j\pb{p_m,\phi_j} \ , \quad \phi_j=0 \ .
 \nonumber \\
 \end{eqnarray}
 In order to determine Lagrange multipliers $\lambda^j$ we demand that
 the constraints $\phi_j=0$ are preserved during the time evolution of the system. Note that it is clear from the form of the equations of motion written
 above that we have to \emph{firstly} calculate Poisson bracket between
 canonical variables and $\phi_j$ and then we can impose the condition $\phi_j=0$. This is the reason why we write $\phi_j\approx 0$ instead of $\phi_j=0$.
 Now the time evolution of the constraint $\phi_i$ is equal to
 \begin{eqnarray}\label{dotphii}
 \dot{\phi}_i=\frac{\partial \phi_i}{\partial t}+
 \frac{\partial \phi_i}{\partial q^m}\dot{q^m}+
 \frac{\partial \phi_i}{\partial p_m}\dot{p}_m=\nonumber \\
 =\frac{\partial \phi_i}{\partial t}+\pb{\phi_i,H_B}+
 \pb{\phi_i,\phi_j}\lambda^j \ .  \nonumber \\
 \end{eqnarray}
 If we impose the condition that the constraint $\phi_i$ is preserved during
 the time evolution of the system we find that the conditions $\dot{\phi}_i=0$
 provide $J$ equations for $J$ unknown $\lambda^i$. Let us now presume non-degenerative case when $\pb{\phi_i,\phi_j}=\triangle_{ij}$ is non-singular matrix so that it has an inverse $\triangle^{jk} \ , \triangle_{ij}\triangle^{jk}=\delta_i^k$. Then (\ref{dotphii}) can be solved as \begin{eqnarray}
 \lambda^i=-\triangle^{ik}\left(\frac{\partial \phi_k}{\partial t}+\pb{\phi_k,H_B}\right) \ .
 \nonumber \\
 \end{eqnarray}
 As a result we find
 that the time evolution of the
 phase space variables $q^m$ and $p_m$ is governed by equations
 \begin{eqnarray}\label{eqtimecona}
& & \dot{q}^m=
 \pb{q^m,H_B}-\pb{q^m,\phi_i}\triangle^{ij}\pb{\phi_j,H_B}-
 \pb{q^m,\phi_i}\triangle^{ij}\frac{\partial \phi_j}{\partial t} \ ,
 \nonumber \\
& & \dot{p}_m=
 \pb{p_m,H_B}-\pb{p_m,\phi_i}\triangle^{ij}\pb{\phi_j,H_B}-
 \pb{p_m,\phi_i}\triangle^{ij}\frac{\partial \phi_j}{\partial t} \ ,
 \nonumber \\
 & & \phi_j=0 \
 \nonumber \\
 \end{eqnarray}
 that can be written in an equivalent form
  \begin{eqnarray}\label{eqtimecon}
 & & \dot{q}^m=
 \pb{q^m,H_B}_D-
 \pb{q^m,\phi_i}\triangle^{ij}\frac{\partial \phi_j}{\partial t} \ ,
 \nonumber \\
 & & \dot{p}_m=
 \pb{p_m,H_B}_D-
 \pb{p_m,\phi_i}\triangle^{ij}\frac{\partial \phi_j}{\partial t} \ ,
 \nonumber \\
 & & \phi_j=0 \ ,
 \nonumber \\
 \end{eqnarray}
 where we introduced Dirac bracket between two phase space functions defined as
 $\pb{X,Y}_D=\pb{X,Y}-\pb{X,\phi_i}\triangle^{ij}\pb{\phi_j,Y}$.
 \subsection{Secondary time dependent constraints}
 Let us now consider situation when the primary constraints $\phi_j(p,q,t)$
 have weakly vanishing Poisson bracket among themselves. Then the requirement of their preservation during the time development of the system has the form
 \begin{equation}
 \frac{d\phi_i}{dt}=\frac{\partial \phi_i}{\partial t}+
 \pb{\phi_i,H_B}+\lambda^j\pb{\phi_i,\phi_j}\approx
 \frac{\partial \phi_i}{\partial t}+
 \pb{\phi_i,H_B}\equiv \phi_i^{II}(p,q,t)\approx 0 \ ,
 \end{equation}
 where now $\phi_i^{II}(p,q,t)$ are secondary constraints.  Finally we have to ensure the preservation of the secondary constraints which implies
 \begin{equation}
 \frac{d\phi_i^{II}}{dt}=\frac{\partial \phi_i^{II}}{\partial t}
 +\pb{\phi_i^{II},H_B}+\lambda^j\pb{\phi_i^{II},\phi_j}\approx 0 \ .
 \end{equation}
 We will presume that the matrix $\pb{\phi_i^{II},\phi_j}\equiv \triangle_{ij}$ is non-singular
 and hence the equation above can be solved for $\lambda^i$ as
 \begin{equation}
 \lambda^i=-\triangle^{ij}\left(
 \frac{\partial \phi^{II}_j}{\partial t}+\pb{\phi^{II}_j,H_B}\right)
 \end{equation}
 and hence the equation of motion for $p^m,q_m$ have the form
 \begin{eqnarray}
 \frac{dq^m}{dt}=\pb{q^m,H_B}+\lambda^j\pb{q^m,\lambda_j}=
 \pb{q^m,H_B}-\pb{q^m,\phi_i}\triangle^{ij}
 \left(\frac{\partial \phi_j^{II}}{\partial t}+\pb{\phi^{II}_j,H_B}\right)
 \nonumber \\
 \end{eqnarray}
and equivalent one for $p_m$.

Let us outline results of the analysis performed in Appendix. We have shown that in case of the explicit time dependent constraints, either primary or secondary, there are additional terms in the equations of motion for canonical variables which are proportional to explicit time derivative of these constraints. These terms are crucial for the equivalence between Lagrangian and
Hamiltonian equations of motion.

\end{document}